\documentclass
[aps,prl,twocolumn,showpacs,lengthcheck,preprintnumbers]{revtex4}%
\usepackage{amsfonts}
\usepackage{amsmath}
\usepackage{amssymb}
\usepackage{graphicx}%
\providecommand{\U}[1]{\protect\rule{.1in}{.1in}}

\begin{document}
\preprint{ }
\title{Effective one-dimensional dynamics of elongated Bose-Einstein condensates}
\author{A. Mu\~{n}oz Mateo}
\email{ammateo@ull.es}
\author{V. Delgado}
\email{vdelgado@ull.es}
\affiliation{Departamento de F\'{\i}sica Fundamental II, Universidad de La Laguna, La
Laguna, Tenerife, Spain}
\date{22 June 2008}


\begin{abstract}
By using a variational approach in combination with the adiabatic
approximation we derive a new effective 1D equation of motion for the axial
dynamics of elongated condensates. For condensates with vorticity $|q|=0$ or
$1$, this equation coincides with our previous proposal [A. Mu\~{n}oz Mateo
and V. Delgado, Phys. Rev. A \textbf{77}, 013617 (2008)]. We also rederive the
nonpolynomial Schr\"{o}dinger equation (NPSE) in terms of the adiabatic
approximation. This provides a unified treatment for obtaining the different
effective equations and allows appreciating clearly the differences and
similarities between the various proposals. We also obtain an expression for
the axial healing length of cigar-shaped condensates and show that, in the
local density approximation and in units of the axial oscillator length, it
coincides with the inverse of the condensate axial half-length. From this
result it immediately follows the necessary condition for the validity of the
local density approximation. Finally, we obtain analytical formulas that give
the frequency of the axial breathing mode with accuracy better than $1\%$.
These formulas can be relevant from an experimental point of view since they
can be expressed in terms only of the axial half-length and remain valid in
the crossover between the Thomas-Fermi and the quasi-1D mean-field regimes. We
have corroborated the validity of our results by numerically solving the full
3D Gross-Pitaevskii equation.

\end{abstract}
\maketitle

\preprint{Annals of Physics \textbf{324}, 709 (2009)}


\section{I. INTRODUCTION}

In recent years there has been great interest in the physics of Bose-Einstein
condensates of dilute atomic gases confined in highly elongated traps
\cite{Olsha1,Petrov1,Dunj1,Das1,Kett1,Strin1,Guerin1,Alex2,RCG}. These systems
are routinely produced experimentally by using microfabricated atom chips
\cite{Hans1,Ott1,Lean3} or tight optical lattices \cite{Greiner1,Moritz1} and
have important applications in generation and manipulation of matter-wave
solitons \cite{Burger1,Weller1} and in the design of highly sensitive quantum
devices such as matter-wave interferometers \cite{Shin2,Schumm1,Wang1}. From a
theoretical point of view, in the mean-field regime and zero temperature
limit, they are accurately described in terms of a macroscopic wave function
$\psi(\mathbf{r},t)$ that satisfies the Gross-Pitaevskii equation (GPE)
\cite{GPE}
\begin{equation}
i\hbar\frac{\partial\psi}{\partial t}=\left(  -\frac{\hbar^{2}}{2m}\nabla
^{2}+V(\mathbf{r})+gN\left\vert \psi\right\vert ^{2}\right)  \psi, \label{I-1}%
\end{equation}
where $N$ is the number of atoms, $g=4\pi\hbar^{2}a/m$ is the interaction
strength determined by the \textsl{s}-wave scattering length $a$, and
$V(\mathbf{r})$ is the potential of the confining trap. In this work we shall
restrict ourselves to condensates with repulsive interatomic interactions
($a>0$).

In highly anisotropic cigar-shaped traps the radial confinement can be so
tight that the transversal motion becomes practically reduced to zero-point
oscillations. Under these circumstances, only the slow axial degrees of
freedom are relevant and the condensate dynamics becomes effectively
one-dimensional. This is the quasi-1D mean-field regime. More generally, a
high anisotropy always induces two very different time scales. When the axial
motion is sufficiently slow in space and time that the radial degrees of
freedom can adjust practically instantaneously to the different axial
configurations, the radial motion becomes irrelevant and one can still study
the condensate dynamics in terms of an effective equation of lower
dimensionality. Several proposals have been made in recent years in this
respect \cite{Jack1,Chio1,Reatto1,Modug1,Kam1,You1}.

By using the adiabatic approximation and the local chemical potential that
follows from a suitable ansatz for the condensate local density \cite{Previos}%
, we derived in Ref. \cite{Anterior} an effective 1D equation of motion for
the axial dynamics of cigar-shaped condensates with repulsive interatomic
interactions. We demonstrated that this equation, which is also applicable to
condensates containing an axisymmetric vortex of topological charge $q$, is
more accurate than previous proposals. It also has the advantage that it
allows to obtain very accurate analytical expressions for a number of
ground-state properties, and these expressions remain valid for condensates
with an arbitrary number of particles. Despite these merits, from a
fundamental point of view it would be desirable to find a more systematic way
of deriving this equation. In this work, by using a variational approach in
combination with the adiabatic approximation we derive a new effective
equation of motion which, for $|q|=0$ and $1$, coincides exactly with our
previous proposal. We also rederive the nonpolynomial Schr\"{o}dinger equation
(NPSE) in terms of the adiabatic approximation. This provides a unified method
for obtaining the different effective equations and allows appreciating
clearly the differences and similarities between the various proposals.
Interestingly, it also demonstrates that in certain cases a variational
approach based on the chemical-potential functional can produce more simple
and accurate results than the usual variational approach based on the energy
functional. We also obtain an expression for the \emph{axial healing length}
of cigar-shaped condensates and show that, in the local density approximation
and in units of the axial oscillator length, it coincides with the inverse of
the axial half-length. From this result it immediately follows the necessary
condition for the validity of the local density approximation. Finally, we
obtain analytic formulas that give the frequency of the axial breathing mode
of an elongated condensate with accuracy better than $1\%$ and remain valid in
the crossover between the Thomas-Fermi (TF) and the quasi-1D mean-field regimes.

\section{II. EFFECTIVE 1D EQUATION OF MOTION}

In Ref. \cite{Anterior} we demonstrated that under usual conditions the axial
dynamics of highly elongated mean-field condensates with repulsive interatomic
interactions, confined in the radial direction by a harmonic potential and
containing, in general, an axisymmetric vortex of charge $q$, can be described
by the effective 1D equation%
\begin{equation}
i\hbar\frac{\partial\phi}{\partial t}=-\frac{\hbar^{2}}{2m}\frac{\partial
^{2}\phi}{\partial z^{2}}+V_{z}(z)\phi+\hbar\omega_{\bot}\sqrt{\beta_{q}%
^{2}+4aN\left\vert \phi\right\vert ^{2}}\phi. \label{II-1}%
\end{equation}
This equation incorporates properly the contribution from the transverse
degrees of freedom through the term proportional to $\hbar\omega_{\bot}$. The
contribution from the vortex is contained entirely in the parameter%
\begin{equation}
\beta_{q}=\frac{2^{2|q|}(|q|!)^{2}}{(2|q|)!}. \label{II-2}%
\end{equation}
The absence of vortices is a particular case corresponding to $q=0$ and
$\beta_{q}=1$. Since vortices with $q\geq2$ are dynamically unstable
\cite{Shin1,PRL06,Mott,Lundh1}, for condensates containing a multiply
quantized vortex the equation above is only applicable up to times shorter
than the vortex decay time.

Equation (\ref{II-1}) was derived in Ref. \cite{Anterior} by applying the
\emph{adiabatic approximation} and using for the corresponding local chemical
potential the analytical expression%
\begin{equation}
\mu_{\bot}(n_{1})=\hbar\omega_{\bot}(|q|+1-\beta_{q})+\hbar\omega_{\bot}%
\sqrt{\beta_{q}^{2}+4an_{1}}. \label{II-3}%
\end{equation}
This expression, in turn, follows from an approximation scheme based on a
suited TF-like ansatz for the condensate local density which, essentially,
represents a simple extension of the Thomas-Fermi approximation \cite{Previos}%
. In what follows we will give an alternative derivation that permits
obtaining the above result in a more systematic way. To this end, it is
convenient to recall very briefly the main steps that led us to Eq.
(\ref{II-1}). Under usual conditions, the time scales characterizing the axial
and the radial dynamics of highly elongated condensates are so different that
one can appeal to the adiabatic approximation and factorize the condensate
wave function as \cite{Jack1,Kramer1}%
\begin{equation}
\psi(\mathbf{r},t)=\varphi(\mathbf{r}_{\bot};n_{1}(z,t))\phi(z,t),
\label{II-4}%
\end{equation}
where $\mathbf{r}_{\bot}=(x,y)$ and $n_{1}(z,t)$ is the local density per unit
length along $z$%
\begin{equation}
n_{1}(z,t)\equiv N\int d^{2}\mathbf{r}_{\bot}|\psi(\mathbf{r}_{\bot}%
,z,t)|^{2}=N|\phi(z,t)|^{2}. \label{II-5}%
\end{equation}
Substituting Eq. (\ref{II-4}) into the GPE and assuming that the axial density
varies sufficiently slowly in space and time, one obtains \cite{Anterior}%
\begin{equation}
i\hbar\frac{\partial\phi}{\partial t}=-\frac{\hbar^{2}}{2m}\frac{\partial
^{2}\phi}{\partial z^{2}}+V_{z}(z)\phi+\mu_{\bot}(n_{1})\phi, \label{II-6}%
\end{equation}%
\begin{equation}
\left(  -\frac{\hbar^{2}}{2m}\nabla_{\bot}^{2}+V_{\bot}(\mathbf{r}_{\bot
})+gn_{1}\left\vert \varphi\right\vert ^{2}\right)  \varphi=\mu_{\bot}%
(n_{1})\varphi, \label{II-7}%
\end{equation}
where we also have assumed a separable confining potential $V(\mathbf{r}%
)=V_{\bot}(\mathbf{r}_{\bot})+V_{z}(z)$. Equation (\ref{II-6}) shows that the
axial dynamics is affected by the radial degrees of freedom only via the
transverse local chemical potential $\mu_{\bot}(n_{1})$. Equation
(\ref{II-7}), which is a \emph{stationary} GPE, reveals that, at every instant
$t$, the transverse wave function $\varphi$ coincides locally with the
equilibrium wave function of an axially uniform condensate characterized by a
linear density $n_{1}(z,t)$. This result simply reflects the fact that, for
highly elongated condensates, at every instant of time the (fast) transverse
degrees of freedom can adjust instantaneously to the local equilibrium
configuration compatible with the axial configuration of the condensate.

In what follows, we shall consider the confining potential to be axisymmetric
and harmonic in the radial direction, while it remains generic in the axial
direction%
\begin{equation}
V(\mathbf{r})=\frac{1}{2}m\omega_{\bot}^{2}r_{\bot}^{2}+V_{z}(z).
\label{II-7b}%
\end{equation}
Using that $\mu_{\bot}(n_{1})$ as given by Eq. (\ref{II-3}) is an accurate
approximate solution of Eq. (\ref{II-7}), after substituting in Eq.
(\ref{II-6}) and taking into account Eq. (\ref{II-5}), one finally arrives at
Eq. (\ref{II-1}).

Other different effective equations of motion can be obtained by using the
adiabatic approximation in combination with a variational approach. To see
this, it is convenient to consider the axially uniform condensate described by
the transverse equation (\ref{II-7}) as composed of an infinite series of
identical pieces of length $L$, with periodic boundary conditions, and
containing $N$ particles each. The exact solutions of this equation are the
critical points of the energy functional%
\begin{equation}
\frac{E[\varphi]}{N}\equiv\!\!\int\!d^{2}\mathbf{r}_{\bot}\left(
\!\frac{\hbar^{2}}{2m}\left\vert \nabla_{\bot}\varphi\right\vert ^{2}+V_{\bot
}\left\vert \varphi\right\vert ^{2}+\frac{1}{2}gn_{1}\left\vert \varphi
\right\vert ^{4}\!\right)  \!, \label{II-8}%
\end{equation}
where $n_{1}=N/L$ is the density per unit length. Thus, the problem of finding
the eigenfunctions $\varphi$ that satisfy Eq. (\ref{II-7}) is equivalent to
the problem of finding, within the \emph{whole} space of admissible functions,
those functions that make the above energy functional stationary. The
corresponding local chemical potential then follows from the relationship:%
\begin{equation}
\mu_{\bot}(n_{1})=\frac{\partial E[\varphi]}{\partial N}. \label{II-8b}%
\end{equation}
In general, however, both problems are equally complicated, so that, in
practice, one usually has to limit the search for the critical points of
$E[\varphi]$ to a subspace of convenient\ variational trial functions. The
solutions so obtained, in general, no longer satisfy exactly the transverse
equation (\ref{II-7}), and the corresponding energy $E[\varphi]$ and chemical
potential $\mu_{\bot}(n_{1})$ can only be considered as mere estimations of
the\ actual values.

Multiplying by $\varphi^{\ast}$ and integrating on the radial coordinates, Eq.
(\ref{II-7}) leads to the chemical-potential functional%
\begin{equation}
\mu_{\bot}[\varphi]\!\equiv\!\!\int\!d^{2}\mathbf{r}_{\bot}\varphi^{\ast
}\!\left(  \!-\frac{\hbar^{2}}{2m}\nabla_{\bot}^{2}+V_{\bot}(\mathbf{r}_{\bot
})+gn_{1}\left\vert \varphi\right\vert ^{2}\!\right)  \!\varphi. \label{II-9}%
\end{equation}
Since for condensates with repulsive interatomic interactions $\mu_{\bot
}[\varphi]\!$ is bounded from below, an independent estimation for the local
chemical potential of the ground state can be obtained by minimizing directly
the above functional. In the ideal-gas perturbative regime ($g\rightarrow0$)
the system becomes quasi-linear and both estimates coincide. In general,
however, Eqs. (\ref{II-8}) and (\ref{II-9}) lead to different results. When
the search is performed within the whole space of admissible functions,
minimization of the energy functional always yields the correct result.
However, as we shall see, when the search is restricted to a subspace of
variational trial functions (as is usually the case), the direct minimization
of the functional (\ref{II-9}) can lead to a better result for the chemical
potential of the ground state.

To the lowest order in the perturbative regime, the ground-state solution of
Eq. (\ref{II-7}) compatible with an axisymmetric vortex of charge $q$ takes
the form \cite{Previos}%
\begin{equation}
\varphi_{q}(r_{\bot},\theta)=\frac{\exp(iq\theta)}{\sqrt{\pi a_{\bot}^{2}%
|q|!}}(r_{\bot}/a_{\bot})^{|q|}\exp(-r_{\bot}^{2}/2a_{\bot}^{2}),
\label{II-10}%
\end{equation}
where $a_{\bot}=\sqrt{\hbar/m\omega_{\bot}}$ is the oscillator length. It is
then natural to look for the critical points of the energy functional
(\ref{II-8}) within the subspace composed of the above functions with the
substitution $a_{\bot}\rightarrow\Gamma a_{\bot}$, where $\Gamma$ is a
dimensionless variational parameter determining the condensate width
\cite{Baym1,Victor1}. Minimization of Eq. (\ref{II-8}) thus yields%
\begin{equation}
\Gamma=\left(  1+\frac{2an_{1}}{(\left\vert q\right\vert +1)\beta_{q}}\right)
^{1/4}. \label{II-11}%
\end{equation}
Substituting\ in Eq. (\ref{II-8}) one obtains the condensate energy%
\begin{equation}
\frac{E}{N}=\hbar\omega_{\bot}(\left\vert q\right\vert +1)\sqrt{1+\frac
{2an_{1}}{(\left\vert q\right\vert +1)\beta_{q}}}. \label{II-12}%
\end{equation}
Using this expression in Eq. (\ref{II-8b}) one finally finds the desired
chemical potential%
\begin{equation}
\mu_{\bot}=\hbar\omega_{\bot}(\left\vert q\right\vert +1)\frac{1+\frac
{3an_{1}}{(\left\vert q\right\vert +1)\beta_{q}}}{\sqrt{1+\frac{2an_{1}%
}{(\left\vert q\right\vert +1)\beta_{q}}}}. \label{II-13}%
\end{equation}
Substitution of Eq. (\ref{II-13}) into Eq. (\ref{II-6}) then leads to the
following effective 1D equation, to be compared to Eq. (\ref{II-1}):%
\begin{multline}
i\hbar\frac{\partial\phi}{\partial t}=-\frac{\hbar^{2}}{2m}\frac{\partial
^{2}\phi}{\partial z^{2}}+V_{z}(z)\phi+\\
\hbar\omega_{\bot}(\left\vert q\right\vert +1)\frac{1+\frac{3aN\left\vert
\phi\right\vert ^{2}}{(\left\vert q\right\vert +1)\beta_{q}}}{\sqrt
{1+\frac{2aN\left\vert \phi\right\vert ^{2}}{(\left\vert q\right\vert
+1)\beta_{q}}}}\phi. \label{II-14}%
\end{multline}
This equation, derived here by using the standard adiabatic approximation, is
nothing but the nonpolynomial Schr\"{o}dinger equation (NPSE) \cite{Salas2}.
Equation (\ref{II-14}) is also applicable to condensates with attractive
interatomic interactions.

As already said, one can still obtain a different effective equation of motion
by estimating the chemical potential directly from the functional
(\ref{II-9}). In doing so, one finds a condensate width%
\begin{equation}
\Gamma=\left(  1+\frac{4an_{1}}{(\left\vert q\right\vert +1)\beta_{q}}\right)
^{1/4},\label{II-15}%
\end{equation}
and, after substitution in Eq. (\ref{II-9}) one arrives at the following
expression for the local chemical potential:%
\begin{equation}
\mu_{\bot}=\hbar\omega_{\bot}(\left\vert q\right\vert +1)\sqrt{1+\frac
{4an_{1}}{(\left\vert q\right\vert +1)\beta_{q}}}.\label{II-16}%
\end{equation}
Substituting again in Eq. (\ref{II-6}) one finally obtains%
\begin{multline}
i\hbar\frac{\partial\phi}{\partial t}=-\frac{\hbar^{2}}{2m}\frac{\partial
^{2}\phi}{\partial z^{2}}+V_{z}(z)\phi+\\
\hbar\omega_{\bot}(\left\vert q\right\vert +1)\sqrt{1+\frac{4aN\left\vert
\phi\right\vert ^{2}}{(\left\vert q\right\vert +1)\beta_{q}}}\phi
.\label{II-17}%
\end{multline}
This equation, to be compared to Eqs. (\ref{II-1}) and (\ref{II-14}), is a new
effective 1D equation governing the axial dynamics of the condensate. As can
be easily verified, for $\left\vert q\right\vert =0$ and $1$ the equation
above coincides exactly with our previous proposal (\ref{II-1}). This is
remarkable since the chemical potentials that lead to both equations have been
derived by applying very different techniques: while the chemical potential
(\ref{II-16}) follows from a variational approach, the chemical potential
(\ref{II-3}) was derived in Ref. \cite{Previos} by using a suited\ TF-like
ansatz for the condensate local density. This nontrivial coincidence provides
additional support to our previous results. Note also that, from a dynamical
point of view, these cases (corresponding to $\left\vert q\right\vert =0$ and
$1$) are precisely the most relevant ones, since for condensates with a larger
vortex charge the applicability of the above equations is limited to times
shorter than the vortex lifetime. For $\left\vert q\right\vert \geq2$, Eqs.
(\ref{II-1}) and (\ref{II-17}) give somewhat different results, which is a
consequence of the different way in which the chemical potentials (\ref{II-3})
and (\ref{II-16}) incorporate the effect of the vortex. The point is to
determine which of the above equations give better results. In Ref.
\cite{Anterior} we demonstrated that, for $q=0$, Eq. (\ref{II-1}) is somewhat
more accurate than Eq. (\ref{II-14}). It remains to be seen whether this is
also the case for condensates containing a vortex. It is clear that the
ability of the above equations for reproducing accurately the axial dynamics
of the condensate is directly related to the ability of the corresponding
chemical potentials (\ref{II-3}), (\ref{II-13}), and (\ref{II-16}) for
reproducing accurately the lowest eigenvalue of the transverse equation
(\ref{II-7}).%
\begin{figure}
[ptb]
\begin{center}
\includegraphics[
height=7.3104cm,
width=8.4526cm
]%
{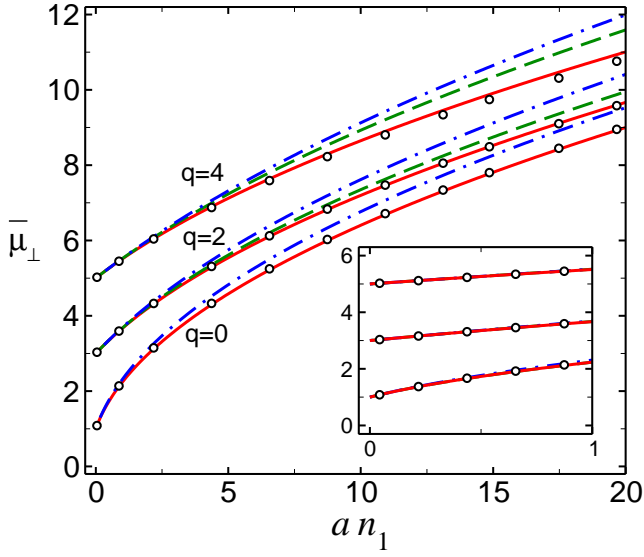}%
\caption{Different theoretical estimates for the local chemical potential
$\overline{\mu}_{\bot}=\mu_{\bot}/\hbar\omega_{\bot}$ as a function of
$an_{1}$. Solid lines have been obtained from Eq. (\ref{II-3}), dashed lines
from Eq. (\ref{II-16}), and dash-dotted lines from Eq. (\ref{II-13}).\ Open
circles are exact results obtained by solving numerically Eq.(\ref{II-7}) with
no approximations.}%
\label{Fig1}%
\end{center}
\end{figure}

Figure \ref{Fig1} compares the different theoretical estimates for the local
chemical potential, obtained from the above formulas, with exact
results\ (open circles) obtained by solving numerically Eq. (\ref{II-7}), with
no approximations, for a wave function of the form $\exp(iq\theta
)\varphi(r_{\bot})$ with $q=0,2$ and $4$. As seen in Ref. \cite{Anterior} our
chemical potential, Eq. (\ref{II-3}), (solid lines) is in \ good agreement
with the numerical results. The maximum error is smaller than $1\%$ for $q=0$
and $2$, and smaller than $2.5\%$ for $q=4$, and this is so for any value of
the dimensionless interaction parameter $an_{1}$ (not only in the range shown
in the figure). As mentioned before, the theoretical estimate from Eq.
(\ref{II-16}) (dashed lines) coincides exactly with that from Eq. (\ref{II-3})
for $q=0$ (and also for $q=1$, not shown in the figure). For $q=2$ the maximum
error (in the range of the figure) is of the order of $3\%$ and for $q=4$ it
is of the order of $7\%$. The estimate from Eq. (\ref{II-13}) (dash-dotted
lines) turns out to be somewhat less accurate. In this case the maximum error
is of the order of $5\%$ for $q=0$, of the order of $7\%$ for $q=2$, and of
the order of $10\%$ for $q=4$. Moreover, these errors continue increasing with
$an_{1}$. These results indicate that the effective 1D equation (\ref{II-1})
should give a better description of the condensate dynamics than the two other
alternative equations.%
\begin{figure}
[ptb]
\begin{center}
\includegraphics[
height=9.2785cm,
width=8.3591cm
]%
{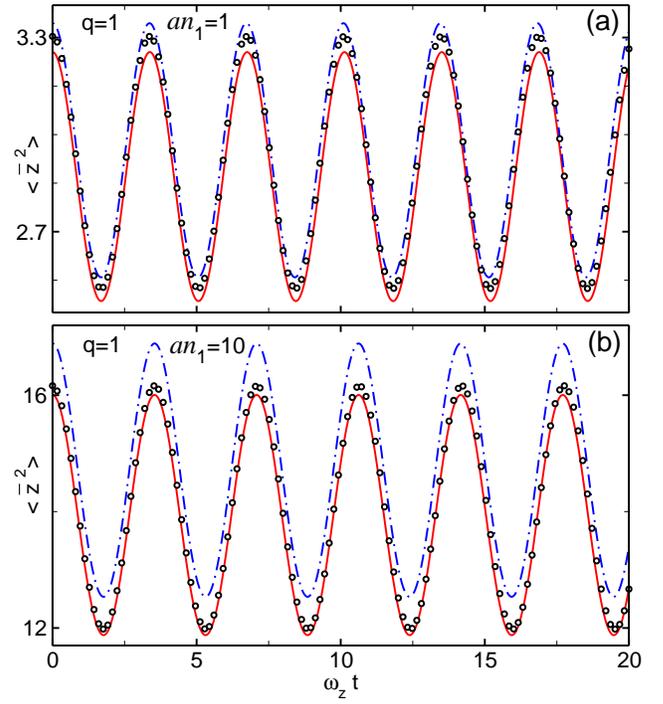}%
\caption{Evolution of $\langle\overline{z}^{2}\rangle$ for an elongated
condensate with $q=1$, after a perturbation that excites its axial breathing
mode ($\overline{z}\equiv z/a_{z}$). Solid lines have been obtained from Eq.
(\ref{II-1}), dashed lines from Eq. (\ref{II-17}), and dash-dotted lines from
Eq. (\ref{II-14}). Open circles are exact results obtained from the full 3D
GPE.}%
\label{Fig2}%
\end{center}
\end{figure}
\begin{figure}
[ptb]
\begin{center}
\includegraphics[
height=9.2785cm,
width=8.3591cm
]%
{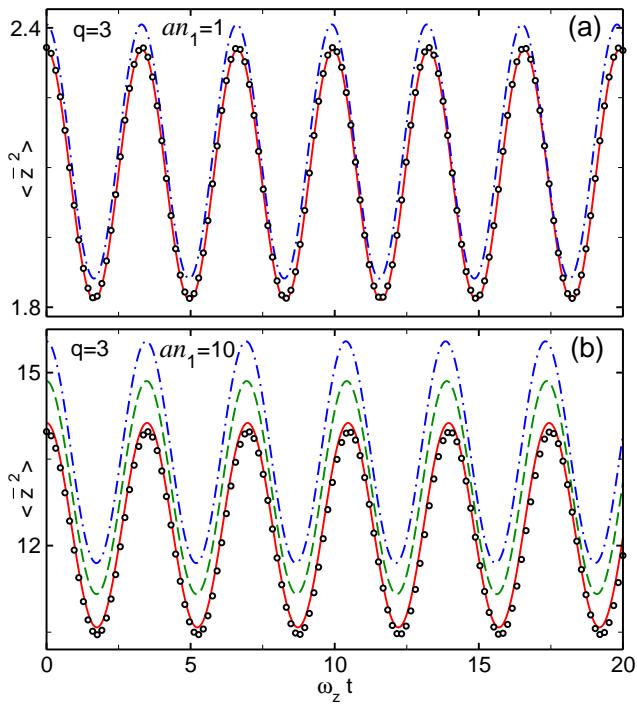}%
\caption{Evolution of $\langle\overline{z}^{2}\rangle$ for an elongated
condensate with $q=3$, after a perturbation that excites its axial breathing
mode ($\overline{z}\equiv z/a_{z}$). Solid lines have been obtained from Eq.
(\ref{II-1}), dashed lines from Eq. (\ref{II-17}), and dash-dotted lines from
Eq. (\ref{II-14}). Open circles are exact results obtained from the full 3D
GPE.}%
\label{Fig3}%
\end{center}
\end{figure}

To verify this point we have studied the dynamical\ evolution of an elongated
condensate with $\lambda=0.1$ after a sudden perturbation that excites its
axial breathing mode. Figures \ref{Fig2} and \ref{Fig3} show the evolution in
time of the mean squared axial amplitude $\langle z^{2}\rangle=N^{-1}\int
dz\,z^{2}n_{1}(z,t)$ after a sudden perturbation at $t=0$ in the axial
confinement frequency of the form $\omega_{z}\rightarrow1.1\omega_{z}$. Figure
\ref{Fig2} corresponds to a condensate containing a vortex of charge $q=1$
while Fig. \ref{Fig3} corresponds to a condensate containing a $q=3$ vortex.
Solid lines have been obtained from our effective equation (\ref{II-1}),
dashed lines have been obtained from the variational equation (\ref{II-17}),
and dash-dotted lines from the NPSE (\ref{II-14}). Open circles are exact
results obtained by solving numerically the full 3D GPE with no
approximations. Since all the three effective equations have the correct
perturbative ($an_{1}\ll1$) limit (see inset in Fig. \ref{Fig1}), they turn
out to be practically indistinguishable in this regime. As Fig. \ref{Fig2}(a)
shows, for condensates with $q=1$ and a (dimensionless) peak axial density
$an_{1}=1$ these equations still give rather similar results. The differences
increase as $an_{1}$ does, as expected from Fig. \ref{Fig1}. This can be
appreciated from Fig. \ref{Fig2}(b) which corresponds to a condensate with
$an_{1}=10$. In this case Eqs. (\ref{II-1}) and (\ref{II-17}), which for $q=1$
are indistinguishable, give more accurate results (solid lines) than Eq.
(\ref{II-14}) (dash-dotted lines).

As Fig. \ref{Fig3}(a) shows, for $q=3$ Eq. (\ref{II-1}) gives somewhat better
results than the two other ones even for condensates with a peak axial density
$an_{1}=1$. In this case, the results from Eq. (\ref{II-17}) (not shown in the
figure for clarity) lie exactly in between the two curves shown. Again, as
Fig. \ref{Fig3}(b) reflects, the differences become more evident as $an_{1}$
increases. These results indicate that the effective 1D equation (\ref{II-1})
is the one that gives a more accurate description of the condensate dynamics.
It is important to note, however, that for condensates with $\left\vert
q\right\vert =0$ or $1$ this equation coincides exactly with the variational
equation (\ref{II-17}).

Equation (\ref{II-1}) has the additional advantage that, in combination with
the local density approximation (whose conditions of validity will be analyzed
in detail in Section III), it allows to derive a number of useful analytical
expressions for the ground-state properties of elongated condensates
\cite{Anterior}. We next give those that will be needed later on. Consider a
trapping potential that is also harmonic in the axial direction $V_{z}%
(z)=\frac{1}{2}m\omega_{z}^{2}z^{2}$, with $a_{z}=\sqrt{\hbar/m\omega_{z}}$
being the axial oscillator length and $\lambda=\omega_{z}/\omega_{\bot}$ the
trap aspect ratio. Then, it can be shown that the total chemical potential is
given by \cite{Anterior}%
\begin{equation}
\frac{\mu}{\hbar\omega_{\bot}}=(|q|+1)+\frac{1}{2}(\sqrt{\lambda}%
\,\overline{Z})^{2}, \label{III-9}%
\end{equation}
where $\overline{Z}=Z/a_{z}$ is the dimensionless axial half-length. The axial
density profile follows from the formula%

\begin{equation}
n_{1}^{0}(z)=\beta_{q}\frac{(\sqrt{\lambda}\;\overline{Z})^{2}}{4a}\left(
1-\frac{z^{2}}{Z^{2}}\right)  +\frac{(\sqrt{\lambda}\;\overline{Z})^{4}}%
{16a}\left(  1-\frac{z^{2}}{Z^{2}}\right)  ^{2} \label{III-10}%
\end{equation}
with $n_{1}^{0}(z)=0$ for $|z|>Z$. The interest of the above formulas lies in
the fact that they are valid and accurate (with an accuracy typically better
than $1\%$) not only in the TF and in the quasi-1D mean-field regime, but also
in between these two limiting cases. Moreover, since they only depend on an
easily measurable physical quantity (the condensate axial half-length) they
can be useful for a more precise experimental characterization of these kind
of systems. In the present work we will use these expressions in order to
derive an approximate analytical formula for the frequencies of the axial
breathing mode of elongated condensates.

It can be shown that the axial half-length satisfies the following polynomial
equation \cite{Anterior}:%
\begin{equation}
\frac{1}{15}(\sqrt{\lambda}\,\overline{Z})^{5}+\frac{1}{3}\beta_{q}%
(\sqrt{\lambda}\,\overline{Z})^{3}=\chi_{1}, \label{III-11}%
\end{equation}
with
\begin{equation}
\chi_{1}\equiv\frac{\lambda Na}{a_{\bot}}. \label{III-11c}%
\end{equation}
An approximate solution that satisfies the above equation for any $\chi_{1}%
\in\lbrack0,\infty)$, with a residual error \cite{Previos} smaller than
$0.75\%$ for $q=0$\ and smaller than $3.2\%$ for $1\leq|q|\leq10$ is given by%
\begin{equation}
\sqrt{\lambda}\,\overline{Z}=\left[  \frac{1}{\left(  15\chi_{1}\right)
^{\frac{4}{5}}+\frac{1}{3}}+\frac{1}{57\chi_{1}+345}+\frac{1}{(3\chi_{1}%
/\beta_{q})^{\frac{4}{3}}}\right]  ^{-\frac{1}{4}} \label{III-12}%
\end{equation}

It is clear that the variational equation (\ref{II-17}) also permits deriving
analytical expressions for the relevant ground-state properties. For $|q|=0$
and $1$ one obtains the same expressions (\ref{III-9})--(\ref{III-12}). For
$|q|\geq2$, the only modification is that the second term on the right-hand
side of Eq. (\ref{III-10}) and the first term on the left-hand side of Eq.
(\ref{III-11}) become multiplied by the factor $\beta_{q}/(|q|+1)$. As
expected from Fig. \ref{Fig1}, these modified equations turn out to be
somewhat less accurate than the above equations, and for this reason, except
for the present Section, we shall not consider them further in this work.
Substitution of Eqs. (\ref{II-5}) and (\ref{II-10}) (with $a_{\bot}%
\rightarrow\Gamma a_{\bot}$) into Eq. (\ref{II-4}) leads to the equilibrium
variational wave function%
\begin{equation}
\psi(\mathbf{r})=\frac{\exp(iq\theta)}{\Gamma a_{\bot}\sqrt{\pi|q|!}}\left(
\frac{r_{\bot}}{\Gamma a_{\bot}}\right)  ^{|q|}\exp(-\frac{r_{\bot}^{2}%
}{2\Gamma^{2}a_{\bot}^{2}})\sqrt{\frac{n_{1}^{0}(z)}{N}}, \label{III-A1}%
\end{equation}
where $\Gamma(n_{1}^{0})$ is the ($z$-dependent) equilibrium\ condensate width
given by Eq. (\ref{II-15}) and $n_{1}^{0}(z)$ is the axial density profile
given by Eq. (\ref{III-10}) (conveniently modified if $|q|\geq2$). On the
other hand, from the equilibrium condensate density $n(\mathbf{r}%
)=N|\psi(\mathbf{r})|^{2}$ one finds the following expression relating the
peak density $n(\mathbf{0})$ with the peak axial density $n_{1}^{0}(0)$ of a
cigar-shaped condensate with no vortices:%
\begin{equation}
n(\mathbf{0})=\frac{n_{1}^{0}(0)}{\pi a_{\bot}^{2}\sqrt{1+4an_{1}^{0}(0)}}.
\label{III-A2}%
\end{equation}
Since $an_{1}^{0}(0)=(\sqrt{\lambda}\;\overline{Z}/2)^{2}+(\sqrt{\lambda
}\;\overline{Z}/2)^{4}$, the above formula can also be used to obtain
$n(\mathbf{0})$ as a function of $Z$. Equation (\ref{III-A2}) has the correct
limits in the two extreme regimes. In particular, in the perturbative regime
($an_{1}\ll1$) it reduces to $n(\mathbf{0})=n_{1}^{0}(0)/\pi a_{\bot}^{2}$,
whereas in the TF regime ($an_{1}\gg1$) it takes the form $n(\mathbf{0}%
)=\sqrt{n_{1}^{0}(0)/a}/2\pi a_{\bot}^{2}$.

\section{III. AXIAL HEALING LENGTH}

For later purposes, it is convenient to rewrite the effective 1D equation
(\ref{II-1}) in terms of an equivalent system of hydrodynamic equations
describing the superfluid dynamics of the condensate \cite{Strin2}. To this
end we write the axial wave function in polar form%
\begin{equation}
\sqrt{N}\phi(z,t)=\sqrt{n_{1}(z,t)}e^{iS(z,t)}. \label{III-1}%
\end{equation}
By substituting in Eq. (\ref{II-1}), one arrives after some algebra at the
following equations governing the evolution in time of the axial density
$n_{1}$ and velocity field $v\equiv(\hbar/m)\partial S/\partial z$:%
\begin{equation}
\frac{\partial n_{1}}{\partial t}+\frac{\partial}{\partial z}(n_{1}v)=0,
\label{III-2}%
\end{equation}%
\begin{equation}
m\frac{\partial v}{\partial t}+\frac{\partial}{\partial z}\!\!\left(
\!\mu_{\bot}\!(n_{1})\!+\!V_{z}\!+\!\frac{1}{2}mv^{2}\!-\!\frac{\hbar^{2}%
}{2m\sqrt{n_{1}}}\frac{\partial^{2}}{\partial z^{2}}\sqrt{n_{1}}\!\right)
\!=\!0. \label{III-3}%
\end{equation}
In these equations, that are completely equivalent to Eq. (\ref{II-1}), the
mean-field interaction energy between atoms enters through the local chemical
potential $\mu_{\bot}(n_{1})$ given by Eq. (\ref{II-3}). The term proportional
to $\hbar^{2}$ is the quantum pressure, which has its origin in the quantum
kinetic energy of the system. Note that $\mu_{\bot}(n_{1})$ can always be
written as%
\begin{equation}
\mu_{\bot}(n_{1})=\hbar\omega_{\bot}(\left\vert q\right\vert +1)+\tilde{\mu
}_{\bot}(n_{1}). \label{III-4}%
\end{equation}
This can be seen from the transverse equation (\ref{II-7}), whose exact
perturbative solution is an infinite series of the form $\mu_{\bot}%
(n_{1})=\hbar\omega_{\bot}(\left\vert q\right\vert +1)+2\beta_{q}^{-1}%
\hbar\omega_{\bot}an_{1}+...$. From Eq. (\ref{III-3}) it is thus clear that
the last term on the right-hand side of Eq. (\ref{III-4}) is the only one that
contributes to the dynamics. Taking this into account, and defining, as usual,
the axial healing length $\xi_{z}$ as the length scale for which the spatial
variations in the condensate density induce a quantum pressure comparable in
magnitude to the contribution from the interaction energy, i.e.,%
\begin{equation}
\frac{\hbar^{2}}{2m\xi_{z}^{2}}\sim\tilde{\mu}_{\bot}(n_{1}), \label{III-5}%
\end{equation}
one obtains the following expression for the axial healing length of elongated
condensates:%
\begin{equation}
\xi_{z}=\frac{\hbar}{\sqrt{2m\left[  \mu_{\bot}(n_{1})-\hbar\omega_{\bot
}(\left\vert q\right\vert +1)\right]  }}. \label{III-6}%
\end{equation}
In the absence of vortices ($q=0$), the above formula reduces to that
introduced in Ref. \cite{Anterior}. When the length scale $\Delta_{z}$ of the
spatial variations of the condensate density along $z$ is much greater than
$\xi_{z}$ the contribution from the quantum pressure becomes negligible in
comparison with the contribution from the interaction energy, and Eq.
(\ref{III-3}) reduces to%
\begin{equation}
m\frac{\partial v}{\partial t}+\frac{\partial}{\partial z}\!\!\left(
\!\mu_{\bot}\!(n_{1})\!+\!V_{z}\!+\!\frac{1}{2}mv^{2}\!\right)  \!=\!0.
\label{III-7}%
\end{equation}
In the stationary state $v=0$, and the above equation leads to%
\begin{equation}
\mu=\mu_{\bot}(n_{1}^{0})+V_{z}(z), \label{III-8}%
\end{equation}
where $n_{1}^{0}(z)$ denotes the equilibrium density per unit length and $\mu$
is the total chemical potential of the condensate. Equation (\ref{III-8}),
which holds whenever $\Delta_{z}\gg\xi_{z}$ and involves no additional
approximations, is the so-called local density approximation.

In what follows, we shall consider a stationary condensate confined by a
trapping potential that is also harmonic in the axial direction $V_{z}%
(z)=\frac{1}{2}m\omega_{z}^{2}z^{2}$. Substituting Eqs. (\ref{III-8}) and
(\ref{III-9}) into Eq. (\ref{III-6}) one finds the following expression for
the \emph{local} axial healing length of an elongated condensate in the local
density approximation:%
\begin{equation}
\frac{\xi_{z}(z)}{a_{z}}=\frac{(Z/a_{z})^{-1}}{\sqrt{1-z^{2}/Z^{2}}}.
\label{III-13}%
\end{equation}
This formula shows that at the condensate edges the axial healing length
diverges, which is an expected result since in this region the kinetic energy
can never be neglected in comparison with the interaction energy. Defining, as
is usual, the healing length of the condensate as the healing length at the
center of the density cloud one arrives at%
\begin{equation}
\frac{\xi_{z}}{a_{z}}=\left(  \frac{Z}{a_{z}}\right)  ^{-1}. \label{III-14}%
\end{equation}
This formula reflects that in the local density approximation and in units of
the axial oscillator length, the healing length of the condensate coincides
with the inverse of its axial half-length. As already said, in order for the
local density approximation to be valid it is sufficient that $\xi_{z}%
\ll\Delta_{z}$. Thus, taking into account that for a condensate in its
stationary state $\Delta_{z}$ is of the order of $Z$, it follows from Eq.
(\ref{III-14}) that the validity of the local density approximation requires
the condition%
\begin{equation}
\frac{Z}{a_{z}}\gg\left(  \frac{Z}{a_{z}}\right)  ^{-1}. \label{III-15}%
\end{equation}
It can be easily verified that in the intermediate and TF regimes the
condition $\lambda\ll1$ (which always holds for elongated condensates) already
guarantees the above requirement. In the quasi-1D mean-field regime, however,
the trap aspect ratio $\lambda$ must be sufficiently small so as to satisfy
the inequality (\ref{III-15}). Taking into account that in this regime
$an_{1}^{0}(0)\simeq\beta_{q}(\sqrt{\lambda}\;\overline{Z}/2)^{2}$, one finds
that the validity of the local density approximation in the perturbative
($an_{1}^{0}\ll1$) regime requires the condition%
\begin{equation}
\lambda\ll\frac{4an_{1}^{0}(0)}{\beta_{q}}. \label{III-16}%
\end{equation}
Alternatively, using that for $\chi_{1}\ll1$ one has $\chi_{1}\simeq(\beta
_{q}/3)(\sqrt{\lambda}\,\overline{Z})^{3}$, the above condition can also be
written as%
\begin{equation}
\lambda\ll\left(  \frac{3\chi_{1}}{\beta_{q}}\right)  ^{2/3}. \label{III-17}%
\end{equation}

\section{IV. COLLECTIVE OSCILLATIONS}

We are interested in obtaining an analytic formula for the frequency of the
axial breathing mode of elongated condensates, applicable also in the
crossover between the TF and the quasi-1D mean-field regimes. To this end we
will use the hydrodynamical equation (\ref{III-7}). Considering small linear
oscillations around the\ equilibrium configuration $n_{1}=n_{1}^{0}+\delta
n_{1}$ and $v=\delta v$, after linearization, Eq. (\ref{III-7}) takes the form%
\begin{equation}
m\frac{\partial}{\partial t}\delta v+\frac{\partial}{\partial z}\left(
\mu_{\bot}(n_{1}^{0})+\left.  \frac{\partial\mu_{\bot}}{\partial n_{1}%
}\right\vert _{0}\delta n_{1}+V_{z}\right)  =0, \label{IV-1}%
\end{equation}
Taking into account Eq. (\ref{III-8}), the above equation reduces to%
\begin{equation}
m\frac{\partial}{\partial t}\delta v+\frac{\partial}{\partial z}\left(
\left.  \frac{\partial\mu_{\bot}}{\partial n_{1}}\right\vert _{0}\delta
n_{1}\right)  =0, \label{IV-2}%
\end{equation}
where, from Eq. (\ref{II-3}), one has%
\begin{equation}
\left.  \frac{\partial\mu_{\bot}}{\partial n_{1}}\right\vert _{0}%
=\frac{2a\hbar\omega_{\bot}}{\sqrt{\beta_{q}^{2}+4an_{1}^{0}}}, \label{IV-3}%
\end{equation}
with $n_{1}^{0}(z)$ given by Eq. (\ref{III-10}). We look for solutions of Eq.
(\ref{IV-2}) in the form of (axial) dilatations \cite{Kagan1,Castin1}, that
is, we introduce a scaling parameter $b(t)$ satisfying the relationships
$z=b(t)z_{0}$ and $v=\dot{b}(t)z_{0}=z\dot{b}(t)/b(t)$, and assume that the
axial density cloud at any $t$ is related to its initial value as%
\begin{equation}
n_{1}(z,t)=\frac{1}{b(t)}n_{1}^{0}(z_{0})=\frac{1}{b(t)}n_{1}^{0}(\frac
{z}{b(t)}). \label{IV-4}%
\end{equation}
For linear oscillations, the case we are interested in, $b(t)=1+\delta b(t)$
and $\delta v=\delta\dot{b}(t)z$ with $\delta b(t)=\delta b_{0}e^{-i\omega t}%
$. Thus, the first term in Eq. (\ref{IV-2}) takes the form%
\begin{equation}
m\frac{\partial}{\partial t}\delta v=-m\omega^{2}z\delta b. \label{IV-5}%
\end{equation}
As for the second term, we note that taking into account Eq. (\ref{IV-4}),
$\delta n_{1}$ can be written as%
\begin{equation}
\delta n_{1}=\left.  \frac{\partial n_{1}}{\partial b}\right\vert _{b=1}\delta
b=-\left[  \left(  1+z\frac{\partial}{\partial z}\right)  n_{1}^{0}(z)\right]
\delta b. \label{IV-6}%
\end{equation}
Using now Eq. (\ref{III-10}), one finds%
\begin{equation}
\delta n_{1}\!=\!\!\left[  \frac{(\sqrt{\lambda}\;\overline{Z})^{2}}{2a}%
\frac{z^{2}}{Z^{2}}\!\left\{  \!\beta_{q}\!+\!\frac{(\sqrt{\lambda}%
\;\overline{Z})^{2}}{2}\!\left(  \!1-\frac{z^{2}}{Z^{2}}\!\right)
\!\!\right\}  \!-n_{1}^{0}\right]  \!\delta b \label{IV-7}%
\end{equation}
From Eqs. (\ref{II-3}), (\ref{III-8}) and (\ref{III-9}) one has%
\begin{equation}
\sqrt{\beta_{q}^{2}+4an_{1}^{0}}-\beta_{q}=\frac{(\sqrt{\lambda}\;\overline
{Z})^{2}}{2}\left(  1-\frac{z^{2}}{Z^{2}}\right)  . \label{IV-8}%
\end{equation}
Using this relationship in Eq. (\ref{IV-7}) we obtain%
\begin{equation}
\delta n_{1}=\left[  \frac{(\sqrt{\lambda}\;\overline{Z})^{2}}{2a}\frac{z^{2}%
}{Z^{2}}\sqrt{\beta_{q}^{2}+4an_{1}^{0}}-n_{1}^{0}\right]  \delta b.
\label{IV-9}%
\end{equation}
Substituting now Eqs. (\ref{IV-3}), (\ref{IV-9}) and (\ref{III-10}) into the
second term of Eq. (\ref{IV-2}), after some algebra one finally obtains%
\begin{equation}
\frac{\partial}{\partial z}\left(  \left.  \frac{\partial\mu_{\bot}}{\partial
n_{1}}\right\vert _{0}\delta n_{1}\right)  =m\Omega^{2}(z)\omega_{z}%
^{2}z\delta b, \label{IV-10}%
\end{equation}
where%
\begin{equation}
\Omega^{2}(z)\equiv\left(  3-\frac{2an_{1}^{0}(z)}{\beta_{q}^{2}+4an_{1}%
^{0}(z)}\right)  . \label{IV-11}%
\end{equation}
From Eqs. (\ref{IV-5}) and (\ref{IV-10}) it is then clear that only in the TF
and the quasi-1D mean-field regimes Eq. (\ref{IV-2}) admit solutions in the
form of dilatations. In the TF regime one has $4an_{1}^{0}\gg\beta_{q}^{2}$
and as a consequence $\Omega^{2}(z)=5/2$. In this case, Eq. (\ref{IV-2})
reflects that the axial breathing mode is an oscillatory dilatation with a
frequency $\omega^{2}=(5/2)\omega_{z}^{2}$, in good agreement with previous
results \cite{Strin2}. In the quasi-1D mean-field regime $4an_{1}^{0}\ll
\beta_{q}^{2}$ and one obtains the well-known result $\omega^{2}=3\omega
_{z}^{2}$ \cite{Csor1,Strin3,Ho1}. In between these two limiting cases,
however, the condensate dynamics does not depart too much from a dilatation.
This is a direct consequence of the fact that $\Omega^{2}(z)$ as given by Eq.
(\ref{IV-11}) is a slowly varying function of $z$ that can be well
approximated by its mean value%
\begin{equation}
\Omega^{2}(z)\approx\frac{1}{2Z}\int_{-Z}^{+Z}\left(  3-\frac{2an_{1}^{0}%
}{\beta_{q}^{2}+4an_{1}^{0}}\right)  dz. \label{IV-12}%
\end{equation}
Substituting Eq. (\ref{III-10}) and performing the integration one obtains the
following analytical expression for the frequency of the axial breathing mode
of a highly elongated condensate:%
\begin{equation}
\frac{\omega^{2}}{\omega_{z}^{2}}=\Omega^{2}\approx\frac{5}{2}+\frac
{1}{2\left(  \zeta^{2}+2\right)  }+\frac{\tanh^{-1}\left(  \zeta/\sqrt
{\zeta^{2}+2}\right)  }{\zeta\left(  \zeta^{2}+2\right)  ^{3/2}},
\label{IV-13}%
\end{equation}
where $\zeta\equiv\sqrt{\lambda/\beta_{q}}\;\overline{Z}$. It can be easily
verified that the above formula has the correct limits in both the TF and the
quasi-1D mean-field regimes. According to Eq. (\ref{IV-13}) one can obtain
$\omega^{2}$ experimentally\ from a measurement of the condensate axial
half-length $Z$ in the equilibrium configuration. Alternatively one can also
use Eq. (\ref{III-12}) to express $\omega^{2}/\omega_{z}^{2}$ as a function
only of $\chi_{1}$ and $\beta_{q}$.

One can still derive an independent analytic estimate for $\omega^{2}$ by
using the formula \cite{Strin1}%
\begin{equation}
\omega^{2}=-2\frac{\langle z^{2}\rangle}{d\langle z^{2}\rangle/d\omega_{z}%
^{2}}. \label{IV-14}%
\end{equation}
This formula, which follows from a sum-rule approach, gives an upper bound to
the frequency of the axial breathing mode. Substituting Eq. (\ref{III-10}) in
the mean squared amplitude%
\begin{equation}
\langle z^{2}\rangle=N^{-1}\int dz\,z^{2}n_{1}^{0}(z), \label{IV-15}%
\end{equation}
and carrying out the integration, Eq. (\ref{IV-14}) becomes
\begin{equation}
\frac{\omega^{2}}{\omega_{z}^{2}}=\frac{8\left(  \zeta^{2}+7\right)  }%
{7-\zeta^{2}-\left(  70+14\zeta^{2}\right)  \Sigma}, \label{IV-16}%
\end{equation}
where $\Sigma=\left(  \lambda/\overline{Z}\right)  d\overline{Z}/d\lambda$.
The derivative with respect to $\lambda$ can be conveniently rewritten as%
\begin{equation}
\frac{d\overline{Z}}{d\lambda}=\frac{1}{\lambda\sqrt{\lambda}}\left[  \chi
_{1}\frac{d(\sqrt{\lambda}\;\overline{Z})}{d\chi_{1}}-\frac{1}{2}%
(\sqrt{\lambda}\;\overline{Z})\right]  . \label{IV-17}%
\end{equation}
Now the derivative on the right-hand side can be obtained from Eq.
(\ref{III-11}). In doing so, one finds%
\begin{equation}
\Sigma=\frac{\chi_{1}}{5\chi_{1}-\frac{2}{3}\beta_{q}(\sqrt{\lambda
}\,\overline{Z})^{3}}-\frac{1}{2}. \label{IV-18}%
\end{equation}
Substituting this result in Eq. (\ref{IV-16}) and using again Eq.
(\ref{III-11}) to eliminate $\zeta^{5}$ in favor of $\zeta^{3}$ one finally
obtains%
\begin{equation}
\frac{\omega^{2}}{\omega_{z}^{2}}=\frac{4\beta_{q}^{5/2}\zeta^{3}-15\chi
_{1}(\zeta^{2}+5)}{3\beta_{q}^{5/2}\zeta^{3}-6\chi_{1}(\zeta^{2}+5)},
\label{IV-19}%
\end{equation}
where $\zeta\equiv\sqrt{\lambda/\beta_{q}}\;\overline{Z}$. As before, taking
into account that according to Eq. (\ref{III-11}) $\chi_{1}$ can be rewritten
in terms of $Z$, the above equation shows that $\omega^{2}$ can be obtained
experimentally\ from a measurement of the axial half-length in the equilibrium
configuration. By using Eq. (\ref{III-12}) one can also rewrite $\omega
^{2}/\omega_{z}^{2}$ as a function only of $\chi_{1}$ and $\beta_{q}$.%
\begin{figure}
[ptb]
\begin{center}
\includegraphics[
height=8.9864cm,
width=8.2615cm
]%
{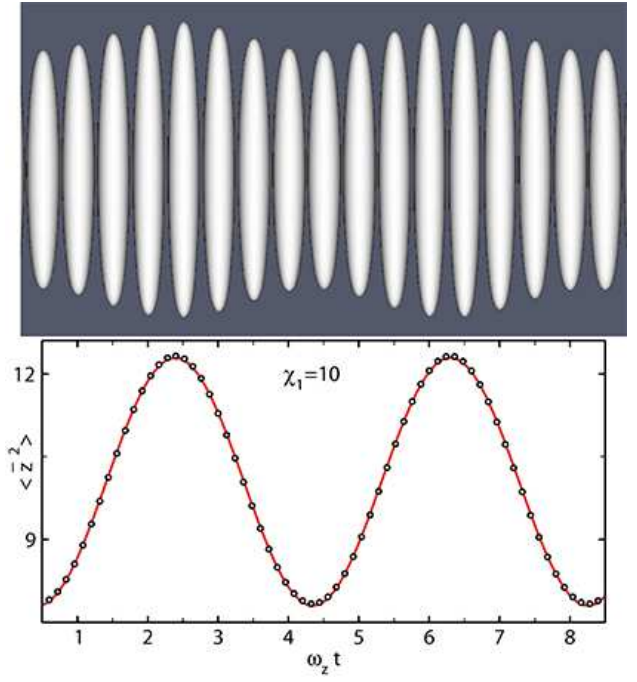}%
\caption{Evolution of a $^{87}$Rb condensate with $\chi_{1}=10$ and no
vortices, in a trap with $\lambda=0.1$ after a $10\%$ fluctuation\ in the
axial trap frequency. Top: isosurfaces of the condensate density taken at
intervals $\Delta t=0.5\omega_{z}^{-1}$. Bottom: evolution of the mean squared
axial amplitude ($\overline{z}\equiv z/a_{z}$). Open circles correspond to the
numerical data and the solid line is the best sinusoidal fit to these data.}%
\label{Fig4}%
\end{center}
\end{figure}
Equations (\ref{IV-13}) and Eq. (\ref{IV-19}) have been obtained under the
assumption that the quantum pressure has a negligible contribution. This
implies that in the quasi-1D mean-field regime ($\chi_{1}\ll1$) the trap
aspect ratio $\lambda$ must be sufficiently small so as to satisfy the
requirement (\ref{III-17}). In the intermediate and TF regimes, however, the
condition $\lambda\ll1$ already guarantees the inequality (\ref{III-15}) to
hold true. It can be easily verified that Eqs. (\ref{IV-13}) and (\ref{IV-19})
coincide each other within $0.65\%$ for $q=0$ and within $0.77\%$ for $q=1$.
In order to determine to what extent the above formulas reproduce the
experimental results we have performed a computer experiment based on the
numerical solution of the full 3D Gross-Pitaevskii equation (\ref{I-1}). We
have considered $^{87}$Rb condensates at zero temperature confined in a
magnetic trap with axial frequency $\omega_{z}=2\pi\times3$ Hz and different
radial frequencies $\omega_{\bot}=\omega_{z}/\lambda$ with $\lambda\leq1/10$.
These condensates have a scattering length $a=5.29$ nm and an oscillator
length $a_{z}=6.23$ $\mu$m. For condensates with a given vortex charge $q$, we
vary the relevant parameter $\chi_{1}=\sqrt{\lambda}Na/a_{z}$ by modifying the
trap aspect ratio $\lambda$ and/or the number of atoms $N$. For a given value
of $\chi_{1}$ we always start with $\lambda=1/10$ and calculate the condensate
equilibrium configuration in the trap. We then excite the axial breathing mode
by introducing a small fluctuation in the axial frequency for a period of time
$t=\omega_{z}^{-1}$ and let the condensate oscillate in the trap. At this
stage we monitor the mean squared axial amplitude as a function of time and
extract the corresponding oscillation frequency from a nonlinear least-squares
fit of a sinusoidal function to the numerical data. We have considered
different perturbations of the trap frequency, ranging from $1\%$ to $10\%$,
and different fitting intervals and have always obtained the same results
within our required precision (better than $0.1\%$). Figure \ref{Fig4}
displays an example corresponding to a condensate with $37220$ atoms
($\chi_{1}=10$) and no vortices, in a trap with an aspect ratio $\lambda
=1/10$. The top panel shows the condensate evolution after a $10\%$
perturbation\ in the axial trap frequency. The images are isosurfaces of the
condensate density (corresponding to $5\%$ of the maximum density) taken at
intervals $\Delta t=0.5\omega_{z}^{-1}$. The bottom panel shows the evolution
of the mean squared axial amplitude in units of $a_{z}^{2}$ ($\overline{z}$
stands for $z/a_{z}$). Open circles correspond to the numerical data and the
solid line is the best sinusoidal fit to these data. From this fit we obtain
an oscillation frequency $\omega=1.605\omega_{z}$ with an accuracy better than
$0.1\%$.

In general, a non-negligible contribution of the quantum pressure manifests
itself as an appreciable dependence of the oscillation frequency on the
parameter $\lambda$. For a given value of $\chi_{1}$, as $\lambda$ increases
the contribution of the quantum pressure also increases, producing in turn an
increase in the condensate oscillation frequency. This fact can be used in
particular to quantify the contribution of the quantum pressure or,
equivalently, the validity of the local density approximation. To guarantee
that the numerical results obtained are comparable with the above analytic
formulas we have repeated the procedure explained before using smaller and
smaller values of $\lambda$ until reaching an oscillation frequency exhibiting
no appreciable dependence on $\lambda$. An estimation of the values required
can be obtained from Eq. (\ref{III-17}).%
\begin{figure}
[ptb]
\begin{center}
\includegraphics[
height=5.934cm,
width=8.7429cm
]%
{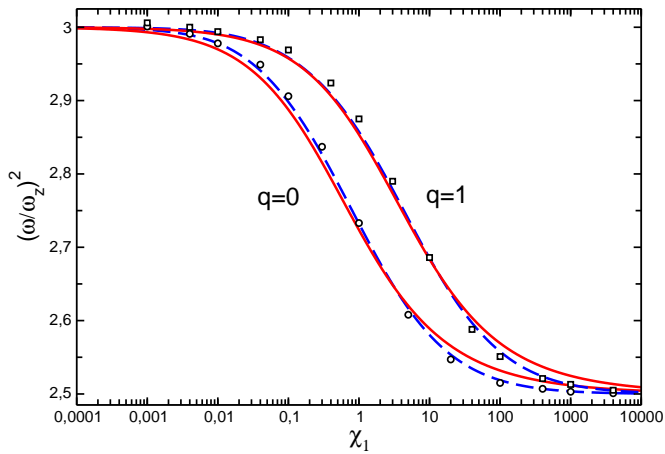}%
\caption{Squared frequency of the axial breathing mode $(\omega/\omega
_{z})^{2}$ of highly elongated condensates with $q=0$ and $1$, as a function
of $\chi_{1}$. Open symbols are numerical results obtained from the full 3D
GPE. Solid curves are the theoretical prediction from Eq. (\ref{IV-13}) and
dashed curves are the theoretical prediction from Eq. (\ref{IV-19}).}%
\label{Fig5}%
\end{center}
\end{figure}
Figure \ref{Fig5} shows the (squared) frequency of the axial breathing mode
$(\omega/\omega_{z})^{2}$ of highly elongated condensates with vorticity $q=0$
and $1$, as a function of $\chi_{1}$. Open symbols are numerical results
obtained from the full 3D GPE with an accuracy of the order of $0.1\%$. Solid
curves are the theoretical prediction from Eqs. (\ref{IV-13}) and
(\ref{III-12}) and dashed curves are the theoretical prediction from Eqs.
(\ref{IV-19}) and (\ref{III-12}). Equation (\ref{IV-13}) reproduces the
numerical results with an accuracy better than $0.8\%$ for $q=0$ and better
than $0.9\%$ for $q=1$. Equation (\ref{IV-19}) is somewhat more accurate. It
reproduces the numerical results with an accuracy better than $0.4\%$ for
$q=0$ and better than $0.65\%$ for $q=1$. As Fig. \ref{Fig5} reflects, in the
crossover between the TF and the quasi-1D mean-field regimes there exists a
clear dependence of the frequency $\omega$ on the vortex charge $q$. This fact
could be used for an indirect detection of unit-charge vortices in the
crossover regime. Because of their accuracy and range of applicability the
above formulas could also be useful from an experimental point of view, for
instance, for calibration of the trap frequencies.

\section{V. CONCLUSION}

In this work, by using the adiabatic approximation in combination with a
variational approach for determining the local chemical potential associated
with the corresponding radial equation, we have derived different effective 1D
equations of motion for the axial dynamics of highly elongated condensates. We
have shown that the minimization of the radial energy functional within a
subspace of convenient\ variational trial functions leads to the NPSE. We also
have demonstrated that in certain cases a variational approach based on the
chemical-potential functional can produce better results than the usual
variational approach based on the energy functional. In this regard, while the
minimization of the energy functional within the whole space of admissible
functions always yields the correct chemical potential, the same does not
remain true when this minimization is restricted to a certain variational
subspace. In fact, we have seen that in\ this case it is possible to obtain a
more simple and yet more accurate result by minimizing directly the
chemical-potential functional. By doing this, we have derived a new effective
1D equation of motion which, for condensates with vorticity $\left\vert
q\right\vert =0$ and $1$, coincides exactly with our previous proposal,
obtained in Ref. \cite{Anterior} by using a suited\ TF-like ansatz. The
variational approach followed in this work provides us with a unified method
for obtaining the different effective 1D equations in a systematic way and
permits us to appreciate clearly the differences and similarities between the
various proposals. A direct comparison with numerical results from the full 3D
GPE indicates that the effective 1D equation proposed in Ref. \cite{Anterior}
is the one that gives a more accurate description of the condensate axial
dynamics. This equation also has the advantage that, in combination with the
local density approximation, it allows to derive accurate analytical formulas
for a number of ground-state properties. These formulas are valid and accurate
(with an accuracy typically better than $1\%$) not only in the TF and in the
quasi-1D mean-field regime, but also in between these two limiting cases.
Moreover, since they only depend on an easily measurable physical quantity
(the condensate axial half-length) they can be useful for a more precise
experimental characterization of these kind of systems.

We also have obtained an expression for the axial healing length of elongated
condensates and have found that, in the local density approximation and in
units of the axial oscillator length, it coincides with the inverse of the
condensate axial half-length. From this result it immediately follows the
necessary condition for the validity of the local density approximation.

Finally, we have obtained approximate analytic formulas that give the
frequency of the axial breathing mode of an elongated condensate with accuracy
better than $1\%$ and remain valid in the crossover between the TF and the
quasi-1D mean-field regimes. These formulas, which as the rest of the
ground-state properties can be expressed in terms only of the axial
half-length $Z$, could be relevant from an experimental point of view since in
the crossover regime (the regime where most experiments are carried out) the
usual formulas are not applicable.

\begin{acknowledgments}
This work has been supported by MEC (Spain) and FEDER fund (EU) (Contract No. Fis2005-02886).
\end{acknowledgments}


\begin{thebibliography}{99}                                                                                               %


\bibitem {Olsha1}M. Olshanii, Phys. Rev. Lett. \textbf{81}, 938 (1998).

\bibitem {Petrov1}D. S. Petrov, G. V. Shlyapnikov, and J. T. M. Walraven,
Phys. Rev. Lett. \textbf{85}, 3745 (2000).

\bibitem {Dunj1}V. Dunjko, V. Lorent, and M. Olshanii, Phys. Rev. Lett.
\textbf{86}, 5413 (2001).

\bibitem {Das1}K. K. Das, Phys. Rev. A \textbf{66}, 053612 (2002).

\bibitem {Kett1}A. G\"{o}rlitz, J. M. Vogels, A. E. Leanhardt, C. Raman, T. L.
Gustavson, J. R. Abo-Shaeer, A. P. Chikkatur, S. Gupta, S. Inouye, T.
Rosenband, and W. Ketterle, Phys. Rev. Lett. \textbf{87}, 130402 (2001).

\bibitem {Strin1}C. Menotti and S. Stringari, Phys. Rev. A \textbf{66}, 043610 (2002).

\bibitem {Guerin1}W. Guerin, J.-F. Riou, J. P. Gaebler, V. Josse, P. Bouyer,
and A. Aspect, Phys. Rev. Lett. \textbf{97}, 200402 (2006).

\bibitem {Alex2}A. I. Nicolin, R. Carretero-Gonz\'{a}lez, and P. G.
Kevrekidis, Phys. Rev. A \textbf{76}, 063609 (2007).

\bibitem {RCG}R. Carretero-Gonz\'{a}lez, D. J. Frantzeskakis and P. G.
Kevrekidis, Nonlinearity \textbf{21}, R139 (2008).

\bibitem {Hans1}W. Hansel, P. Hommelhoff, T.W. Hansch, and J. Reichel, Nature
\textbf{413}, 498 (2001).

\bibitem {Ott1}H. Ott, J. Fortagh, G. Schlotterbeck, A. Grossmann, and C.
Zimmermann, Phys. Rev. Lett. \textbf{87}, 230401 (2001).

\bibitem {Lean3}A. E. Leanhardt, A. P. Chikkatur, D. Kielpinski, Y. Shin, T.
L. Gustavson, W. Ketterle, and D. E. Pritchard, Phys. Rev. Lett. \textbf{89},
040401 (2002).

\bibitem {Greiner1}M. Greiner, I. Bloch, O. Mandel, T. W. H\"{a}nsch, and T.
Esslinger, Phys. Rev. Lett. \textbf{87}, 160405 (2001).

\bibitem {Moritz1}H. Moritz, T. St\"{o}ferle, M. K\"{o}hl, and T. Esslinger,
Phys. Rev. Lett. \textbf{91}, 250402 (2003).

\bibitem {Burger1}S. Burger, K. Bongs, S. Dettmer, W. Ertmer, K. Sengstock, A.
Sanpera, G. V. Shlyapnikov, and M. Lewenstein, Phys. Rev. Lett. \textbf{83},
5198 (1999).

\bibitem {Weller1}A. Weller, J. P. Ronzheimer, C. Gross, J. Esteve, M. K.
Oberthaler, D. J. Frantzeskakis, G. Theocharis, and P. G. Kevrekidis, Phys.
Rev. Lett. \textbf{101}, 130401 (2008).

\bibitem {Shin2}Y. Shin, M. Saba, T. A. Pasquini, W. Ketterle, D. E.
Pritchard, and A. E. Leanhardt, Phys. Rev. Lett. \textbf{92}, 050405 (2004).

\bibitem {Schumm1}T. Schumm, S. Hofferberth, L. M. Andersson, S. Wildermuth,
S. Groth, I. Bar-Joseph, J. Schmiedmayer and P. Kr\"{u}ger, Nat. Phys.
\textbf{1}, 57 (2005).

\bibitem {Wang1}Y.-J. Wang, D. Z. Anderson, V. M. Bright, E. A. Cornell, Q.
Diot, T. Kishimoto, M. Prentiss, R. A. Saravanan, S. R. Segal, and S. Wu,
Phys. Rev. Lett. \textbf{94}, 090405 (2005).

\bibitem {GPE}E. P. Gross, Nuovo Cimento \textbf{20}, 454 (1961); J. Math.
Phys. \textbf{4}, 195 (1963); L. P. Pitaevskii, Zh. Eksp. Teor. Fiz.
\textbf{40}, 646 (1961) [Sov. Phys. JETP \textbf{13}, 451 (1961)].

\bibitem {Jack1}A. D. Jackson, G. M. Kavoulakis, and C. J. Pethick, Phys. Rev.
A \textbf{58}, 2417 (1998).

\bibitem {Chio1}M. L. Chiofalo and M. P. Tosi, Phys. Lett. A \textbf{268}, 406 (2000).

\bibitem {Reatto1}L. Salasnich, A. Parola, and L. Reatto, Phys. Rev. A
\textbf{65}, 043614 (2002).

\bibitem {Modug1}P. Massignan and M. Modugno, Phys. Rev. A \textbf{67}, 023614 (2003).

\bibitem {Kam1}A. M. Kamchatnov and V. S. Shchesnovich, Phys. Rev. A
\textbf{70}, 023604 (2004).

\bibitem {You1}W. Zhang and L. You, Phys. Rev. A \textbf{71}, 025603 (2005).

\bibitem {Previos}A. Mu\~{n}oz Mateo and V. Delgado, Phys. Rev. A \textbf{75},
063610 (2007); Phys. Rev. A \textbf{74}, 065602 (2006).

\bibitem {Anterior}A. Mu\~{n}oz Mateo and V. Delgado, Phys. Rev. A
\textbf{77}, 013617 (2008);

\bibitem {Shin1}Y. Shin, M. Saba, M. Vengalattore, T. A. Pasquini, C. Sanner,
A. E. Leanhardt, M. Prentiss, D. E. Pritchard, and W. Ketterle, Phys. Rev.
Lett. \textbf{93}, 160406 (2004).

\bibitem {PRL06}A. Mu\~{n}oz Mateo and V. Delgado, Phys. Rev. Lett.
\textbf{97}, 180409 (2006).

\bibitem {Mott}J. A. Huhtam\"{a}ki, M. M\"{o}tt\"{o}nen, T. Isoshima, V.
Pietil\"{a}, and S. M. Virtanen, Phys. Rev. Lett. \textbf{97}, 110406 (2006).

\bibitem {Lundh1}H. M. Nilsen and E. Lundh, Phys. Rev. A \textbf{77}, 013604 (2008).

\bibitem {Kramer1}M. Kr\"{a}mer, C. Menotti and M. Modugno, J. Low Temp. Phys.
\textbf{138}, 729 (2005).

\bibitem {Baym1}G. Baym and C. J. Pethick, Phys. Rev. Lett. \textbf{76}, 6 (1996).

\bibitem {Victor1}V. M. P\'{e}rez-Garc\'{\i}a, H. Michinel, J. I. Cirac, M.
Lewenstein, and P. Zoller, Phys. Rev. A \textbf{56}, 1424 (1997).

\bibitem {Salas2}L. Salasnich, B. A. Malomed, and F. Toigo, Phys. Rev. A
\textbf{76}, 063614 (2007).

\bibitem {Strin2}S. Stringari, Phys. Rev. Lett. \textbf{77}, 2360 (1996).

\bibitem {Kagan1}Yu. Kagan, E. L. Surkov, and G. V. Shlyapnikov, Phys. Rev. A
\textbf{54}, R1753 (1996).

\bibitem {Castin1}Y. Castin and R. Dum, Phys. Rev. Lett. \textbf{77}, 5315 (1996).

\bibitem {Csor1}A. Csordas and R. Graham, Phys. Rev. A \textbf{59}, 1477 (1999).

\bibitem {Strin3}S. Stringari, Phys. Rev. A \textbf{58}, 2385 (1998).

\bibitem {Ho1}T.-L. Ho and M. Ma, J. Low Temp. Phys. \textbf{115}, 61 (1999).

\bigskip
\end{thebibliography}
\end{document}